\documentclass{aastex62}
\usepackage{amsmath}
\graphicspath{{./}{figures/}}

\published{08/05/2020}

\shorttitle{First detection of the GI-type of IA using the self-calibration}
\shortauthors{Pedersen et al.}

\begin{document}

\title{First detection of the GI-type of intrinsic alignments of galaxies using the self-calibration 
method in a photometric galaxy survey}

\correspondingauthor{Mustapha Ishak}
\email{mishak@utdallas.edu}

\author{Eske M. Pedersen}
\affil{Department of Physics,
The University of Texas at Dallas,
Richardson, TX 75080, USA}
\author{Ji Yao}
\affiliation{Department of Astronomy,
Shanghai Jiao-Tong University,
Shanghai 200240, China}
\affiliation{Department of Physics,
The University of Texas at Dallas,
Richardson, TX 75080, USA}
\author{Mustapha Ishak}
\affil{Department of Physics,
The University of Texas at Dallas,
Richardson, TX 75080, USA}
\author{Pengjie Zhang}
\affiliation{Department of Astronomy,
Shanghai Jiao-Tong University,
Shanghai 200240, China}



\begin{abstract}

Weak gravitational lensing is one of the most promising cosmological probes to constrain dark matter, dark energy, and the nature of gravity at cosmic scales. Intrinsic alignments (IAs) of galaxies have been recognized as one of the most serious systematic effects facing gravitational lensing. Such alignments must be isolated and removed to obtain a pure lensing signal. Furthermore, the alignments are related to the processes of galaxy formation, so their extracted signal can help in understanding such formation processes and improving their theoretical modeling. We report in this Letter the first detection of the gravitational shear--intrinsic shape (GI) correlation and  the intrinsic shape--galaxy density (Ig) correlation using the self-calibration method in a photometric redshift survey. These direct measurements are made from the KiDS-450 photometric galaxy survey with a significance of 3.65$\sigma$ in the third bin for the Ig correlation, and 3.51$\sigma$ for the GI cross-correlation between the third and fourth bins. The self-calibration method uses the information available from photometric surveys without needing to specify an IA model and will play an important role in validating IA models and IA mitigation in future surveys such as the Rubin Observatory Legacy Survey of Space and Time, Euclid, and WFIRST. 

\end{abstract}



\section{Introduction} \label{sec:intro}
In the past few decades, cosmology has entered a flourishing era of high precision made possible by the advancement of astronomical surveys and missions. 
These will continue to provide large-volume, high-quality observational data that will allow the scientific community to put stringent constraints on cosmological models of the universe. With such an abundance of data, it has become clear that the challenges facing modern cosmology lie in systematic uncertainties associated with the data rather than statistical ones.       
 
One of the most powerful cosmological probes of large-scale structure and matter in the universe is weak gravitational lensing, also known as cosmic shear. Weak gravitational lensing is the physical phenomenon where images of billions of background galaxies are distorted and harmonically aligned by the foreground dark matter and galaxies. These distorted images encode valuable cosmological information about the intervening cosmos that light traveled through. Depending on the position of the sources, lenses, and the observer, gravitational lensing occurs: in a strong regime, giving astonishing multiple images; an in  intermediate regime, giving arcs and arclets; and in a weak regime, giving small distortions of the images of background galaxies. For more details see the reviews \cite{1992grle.book.....S,2015RPPh...78h6901K} and references therein. 
\begin{figure}
\plottwo{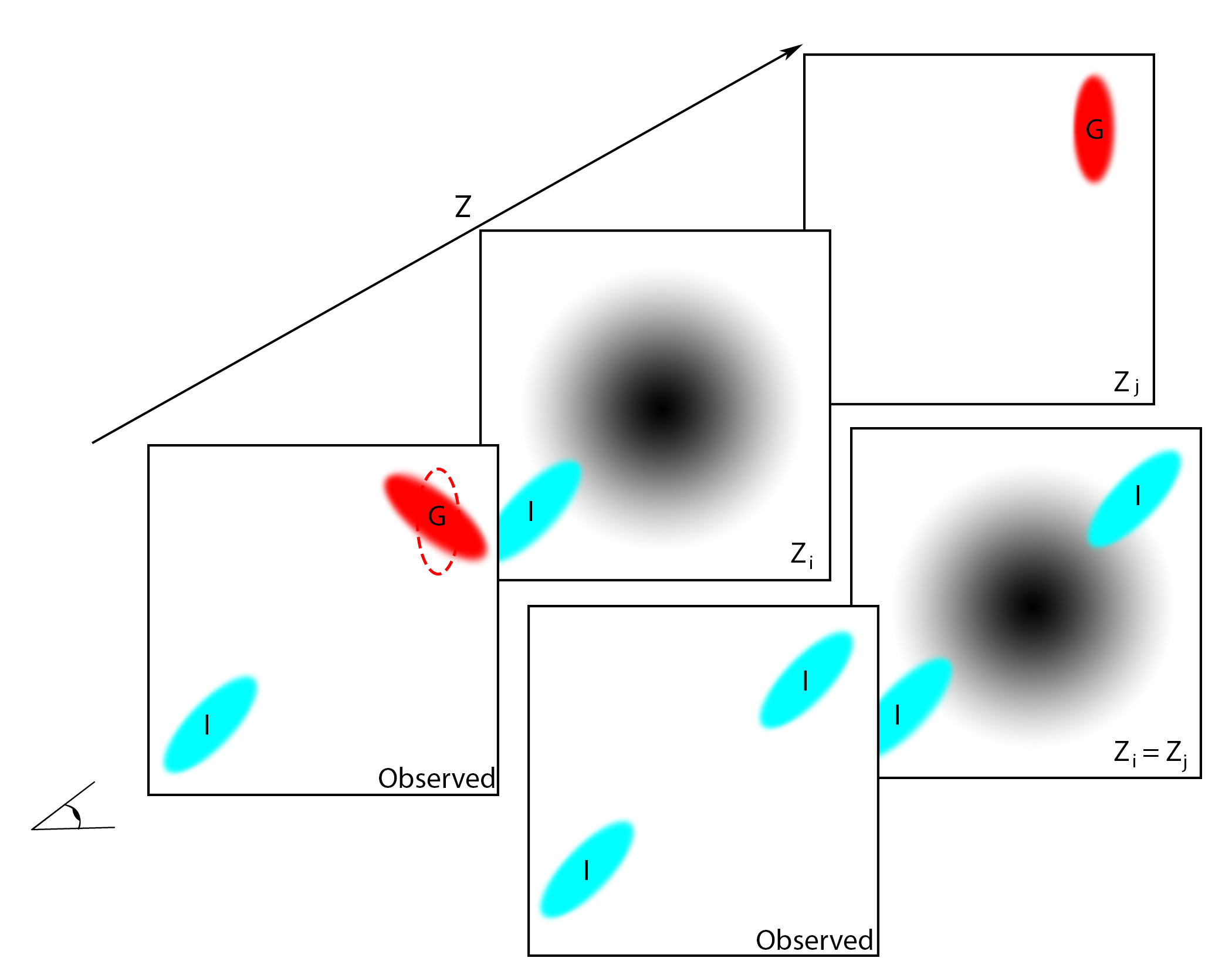}{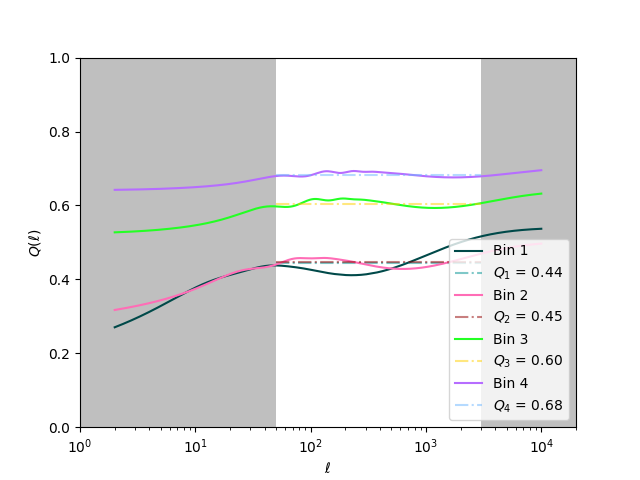}
\caption{Left: a simple illustration of the intrinsic shape--gravitational shear (IG) and intrinsic shape.-intrinsic shape (II) signals. We adopt here the convention that because the I is closer to us it goes first, whereas this relation is often called GI in the literature. The bottom-left-most panels are what is observed: while we see that the light from the (G) galaxy (in red) is getting sheared, i.e. distorted, by the intervening matter, this distortion does not happen to the intrinsically aligned (I) galaxy (in blue). This creates an (anti)correlation IG. In the bottom two panels we see the effect when the two galaxies are around the same redshift and are both aligned toward the same matter-halo creating the II correlation. 
Right: a plot of the $Q(\ell)$ for the four different bins in the KiDS-450 dataset. The $Q$s are calculated using the second pipeline's $Q$ algorithm. Also shown is the averaged $Q$ for the highlighted area, which spans $50\ell$ to $3000\ell$. The $Q$s are reasonably constant for the high-redshift bins, while for the low-redshift bins this is clearly not the case.Throughout this Letter we focus on the two high-redshift bins, since we need $Q$ to be constant.}
\label{fig:IAmodel}
\end{figure}

The effect in the weak regime is tiny but overwhelmingly abundant and is collected by surveys using statistical methods to build a powerful signal to constrain cosmological model parameters. Weak lensing is sensitive to the amount and distribution of matter in the universe as well as the parameters of the dark energy driving the acceleration of the Universe. Weak lensing also probes the growth rate of large-scale structures in the universe which allows it to test the theory of gravity at cosmological scales. A number of weak lensing surveys such as CFHTLens, KiDS-450, and Dark Energy Survey have already delivered -- in combination with other probes -- very tight constraints on the amount of matter, the amplitude of matter clustering, and equation of state of dark energy; see e.g. \cite{2012MNRAS.427..146H,2017MNRAS.465.1454H,2018PhRvD..98d3528T}. Weak lensing is thus found to be one of the most promising cosmological probes, and a number of ambitious surveys are being built and scheduled to start taking data in the upcoming decade, including the Rubin Observatory Legacy Survey of Space and Time (LSST), Euclid, and WFIRST. Again, all these surveys will be dominated by systematic uncertainties, and the scientific community is working on such systematics as uncertainties on photometric redshifts, intrinsic alignments (IAs) of galaxies, baryonic effects, and modeling of nonlinear regimes, among others; for more details see, for example, the reviews \cite{2013MNRAS.429..661M,2018ARA&A..56..393M} and references therein.   

Undoubtedly, one of the most serious systematic effects that weak lensing surveys face is the so-called IAs of galaxies that act as a contaminant to the weak gravitational lensing signal. Galaxies in the universe are not randomly aligned but rather possess an intrinsic alignment due to how they formed and the environment they formed into. More detail can be found in, for example, \cite{2015PhR...558....1T} and references therein. 
Indeed these IAs generate additional signals that contaminate the pure cosmic gravitational shear and significantly affect the values of cosmological parameters; see, e.g. \cite{Schaefer:2015rba}. Studies have shown (e.g. \cite{2007NJPh....9..444B}) that IAs, if not accounted for in weak lensing cosmological analyses, lead to biases (shifts) of up to 30\% in the amplitude parameter of matter fluctuations in the universe and up to 50\% in the equation of state of dark energy. 

To complicate the issue, there are two types of IAs that require different methods of mitigation. First, a collection of galaxies formed around a massive dark matter structure will tend to be radially aligned toward such a structure. This type of IA is called the intrinsic shape-intrinsic shape correlation and is referred to as II. The other type of intrinsic alignment is slightly more subtle and comes from the fact that the same massive matter structure will not only radially align a galaxy close to it but also lenses the image of a background galaxy. This creates an anticorrelation between the images of the two galaxies on the observed sky. This effect is called the gravitational shear--intrinsic shape correlation and is referred to as GI (or IG) signal. The two effects, II and IG, are illustrated in Figure \ref{fig:IAmodel}. 

The scientific community working on weak lensing cosmology and the communities working on preparing software pipelines for upcoming photometric surveys have a strong need for efficient methods to mitigate and control the IA nuisance effect. While the effect of the II signal of IA can be reduced by not considering pairs of galaxies close to each other along the line of sight (i.e. not the same redshift bins), the GI signal cannot be reduced in the same way as it is present at long distances.  
One method used to try to account for GI is to assume a model of IAs with a few parameters and then add those parameters to the cosmological analysis such that the parameters can be constrained from the photometric survey data. This technique relies on the knowledge and specification of an IA model that is still an area of active development itself; see, e.g.,  \cite{2009IJMPD..18..173S,2014MNRAS.445..726C,Leonard2018,Vlah:2019byq}.
Another proposed mitigation method is the nulling technique that uses different redshift dependencies of lensing and IA but it was found to throw away too much of the precious lensing signal \cite{Joachimi:2008ea}. A third scheme that was proposed in \cite{Zhang:2008pw} for the 2-point correlations and later restudied and extended to 3-point correlations in \cite{Troxel:2011za} is called the self-calibration method. As we describe in the next section, we use all the observed correlations between shapes and densities of galaxies in a photometric survey and put them into a procedure that will separate the GG and GI signals. This separation is based on using the dependencies of GG and GI on the respective positions of the sources and lenses in small redshift bins but still allowing the use of the whole redshift extent of the survey. 
Self-calibration in this context means the use of  the data available in the survey itself to calculate a few extra correlations, without the need of an external intrinsic alignment model to obtain an estimate of the data’s contamination by Intrinsic alignment, which in turn can be calibrated (mitigated) out of the data itself.
Ref. \cite{2017JCAP...10..056Y} showed how such a method can mitigate biases on the dark energy parameters. Therefore, self-calibration complements the marginalization method as it does not rely on the specification of an IA model. It allows one to extract the GI signal that can be then subtracted from the GG signal before performing cosmological analyses. Additionally, self-calibration provides the extracted GI signal that can be fit to models of IA and help study and improve such models.  

In this Letter, we report first detections of intrinsic shape--gravitational shear (IG) and intrinsic shape--galaxy density (Ig) in a photometric redshift survey using the self-calibration method where no IA model has been assumed. We provide a concise description of intrinsic alignment, the self-calibration method, the steps that directly led to the detections, and the results obtained. A more detailed description of the technical aspects of the method and the pipelines, the Ig part of the results, and  other developments can be found in a companion paper \cite{2020MNRAS.495.3900Y}.

\section{Intrinsic alignments of galaxies and basic elements of the self-calibration method.} \label{sec:basicselfcalibration}
In photometric galaxy surveys, the total measured shear is given by $\gamma^{obs}=\gamma^G+\gamma^I+\gamma^N$, where the superscript G stands for gravitational shear, I for intrinsic alignment, and N for shot noise. Thus, the observed angular cross-correlation, $<\gamma^{obs,i},\gamma^{obs,j}>$, between two redshift bins $i$ and $j$ includes: a GG term that corresponds to the genuine gravitational shear signal; GI, II, and IG terms that represent intrinsic alignment components; and a noise term. This can be written in terms of the corresponding shape-shape power spectrum as follows:      
\begin{equation}
C^{\gamma\gamma}_{ij}(\ell)=C^{GG}_{ij}(\ell)+C^{IG}_{ij}(\ell)+C^{GI}_{ij}(\ell)+C^{II}_{ij}(\ell)+\delta_{ij}C^{GG,N}_{ii}.
\end{equation} 
Figure \ref{fig:IAmodel} shows the physical mechanisms behind the correlation giving the terms $C^{II}_{ij}(\ell)$ and  $C^{IG}_{ij}(\ell)$. Note that we use here the convention that the IG term represents the intrinsic alignment signal and that the GI term should become negligible.

The components $C^{GI}_{ij}(\ell)$ and $C^{II}_{ij}(\ell)$ can be minimized by choosing bins with $i<j$. $C^{GI}_{ij}(\ell)$ will be minimal due to G being in front of I so no such correlation can be present, while the $C^{II}_{ij}(\ell)$ term is negligible since it is present only for close galaxies but not between galaxies in distinct bins.  
\begin{figure}
\plottwo{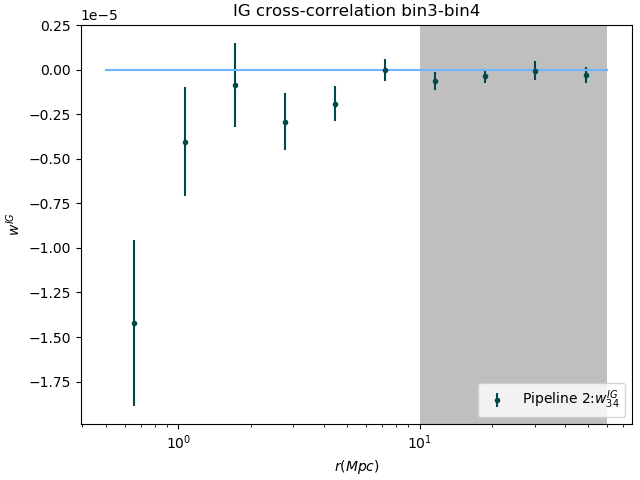}{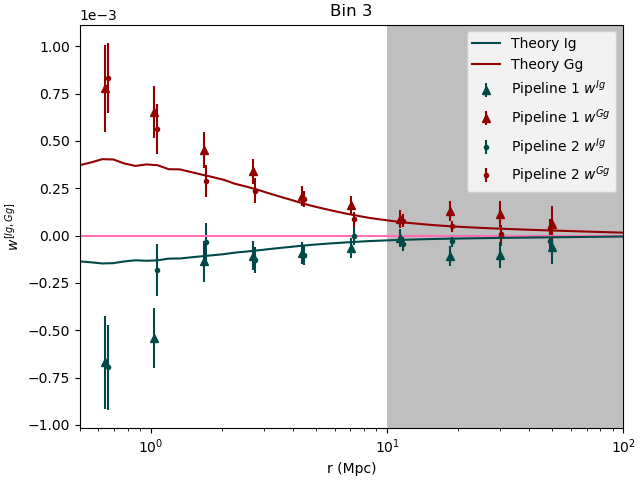}
\caption{\label{fig:IG-signal}
Left: intrinsic shape--gravitational shear (IG) signal from cross-correlating bin 3 and bin 4 from the KiDS-450 dataset. This has been derived using pipeline-2 with a significance of \textcolor{black}{ 3.51$\sigma$}. The gray area is excluded based on theoretically weak expected signal, as well as the breaking of the constant $Q$ approximation. 
Right: intrinsic shape--galaxy density (Ig) signal, and the Gravitational shear--galaxy density (Gg) signal for bin 3, using the two pipelines. Pipeline-1 results are marked with triangles, while pipeline-2's results are marked with dots. Also plotted are a theoretical Gg signal and a theoretical Ig signal, for the best-fit cosmology of KiDS-450, using the default tidal alignment model for the Ig using the best-fit $A_\text{IA}$ amplitude found in \cite{2017MNRAS.465.1454H}. For the second pipeline, we find that the detection of Ig has a $3.65\sigma$ significance. The errors bars are estimated from the jackknife method as described in the text.}
\end{figure}

Now, the self-calibration is used to separate the two remaining terms, $C^{GG}_{ij}(\ell)$ and $C^{IG}_{ij}(\ell)$. First, in the small redshift bin approximation, a scaling relation was derived to relate the IG term to the Ig term \cite{Zhang:2008pw}:

\begin{equation}
    \label{eq:scaling}
    C_{ij}^{IG}\left(\ell\right) \simeq \frac{W_{ij}\Delta_i}{b_i\left(\ell\right)} C_{ii}^{Ig}\left(\ell\right)
\end{equation}

where $W_{ij}$ is the weighted lensing kernel:
\begin{equation}
    W_{ij} = \int_0^\infty \mathrm{d}z_L \int_0^\infty \mathrm{d}z_S W_L\left(z_L,z_S\right) n_i\left(z_L\right)n_j\left(z_S\right),
\end{equation}

with $z_L$ and $z_S$ being the redshift of the lens and source, respectively, and $W_L$ being the lensing kernel given by
\begin{eqnarray}
   W_L\left(z_L, z_S\right) = \begin{cases}
   \frac{3}{2}\Omega_m \frac{H_0^2}{c^2}\left(1+z_L\right) \chi_L \left(1-\frac{\chi_L}{\chi_S}\right) &\text{for} z_L < z_S\\
   0 &\text{otherwise.}
   \end{cases}
\end{eqnarray}

Meanwhile, $\Delta_i$ is the effective width of the $i$th bin:
\begin{equation}
    \Delta_i^{-1} = \int_0^\infty n_i^2\left(z\right) \frac{\mathrm{d}z}{\mathrm{d}\chi}\mathrm{d}z,
\end{equation}
and $b_i$ is the galaxy bias in the $i$th bin.

We use the self-calibration method including the Hankel transform of Equation \ref{eq:scaling} to measure the $w_\text{GI}$ correlation signal in the KiDS-450 dataset. Following the approach outlined in \cite{Zhang:2008pw,Troxel:2011za}, we start by defining the selection function $(S)$, which selects only pairs of galaxies with photometric redshifts $z_G^P < z_g^P$, for the photometric bin. That is,  
\begin{eqnarray}
    \label{selection}
    S(z^P_G,z^P_g)=\begin{cases}
    1 &\text{for $z^P_G<z^P_g$}\\
    0 &\text{otherwise.}
    \end{cases}
\end{eqnarray}

We then build the following two observables:
	\begin{subequations}
		\begin{align} 
		C^{\gamma g}_{ii}&=C^{Ig}_{ii}+C^{Gg}_{ii}, \label{gamma-g} \\
		C^{\gamma g}_{ii}|_S&=C^{Ig}_{ii}+C^{Gg}_{ii}|_S \label{gamma-g|S}.
		\end{align}
	\end{subequations}

Now, since the IA signal does not depend on the ordering of the source--lens pair (contrary to the lensing signal), one can write $C^{Ig}_{ii}|_S=C^{Ig}_{ii}$, while $C^{Gg}_{ii}|_S < C^{Gg}_{ii}$. 
Next, we define the parameter $Q_i$ that quantifies how well we can distinguish the galaxy shear--galaxy density (Gg) signal with or without the selection function:

\begin{equation}
    \label{eq:Q}
    Q_i\left( \ell \right) = \frac{\left.C^{Gg}_{ii}\left(\ell\right)\right|_S}{C^{Gg}_{ii}\left(\ell\right)}
\end{equation}

To calculate this we use the following spectra:

\begin{eqnarray}
    C_{ii}^{Gg}\left(\ell\right) &= \int_0 ^\infty \frac{W_i\left(\chi\right)n_i\left(\chi\right)}{\chi^2}b_g P_\delta \left(k=\frac{\ell}{\chi};\chi\right) \mathrm{d}\chi \label{eq:CGg}\\
    \left.C^{Gg}_{ii}\left(\ell\right)\right|_S &= \int_0 ^\infty \frac{W_i\left(\chi\right)n_i\left(\chi\right)}{\chi^2}b_g P_\delta \left(k=\frac{\ell}{\chi};\chi\right) \eta_i\left(z\right)\mathrm{d}\chi,\label{eq:CGgS}
\end{eqnarray}
where $W_i$ is the lensing efficiency, $n_i$ is the true redshift distribution, $\chi$ is the comoving distance, \textcolor{black}{$b_g$ is the galaxy bias that is assumed to be approximately constant over the bin (this is found to be the case from our galaxy bias explicit calculation for the four bins)}, $P_\delta$ is the matter power-spectra, and $\eta_i$ is a function of the selection function that was defined in \cite{Zhang:2008pw} as

\begin{equation}
\eta_i(z) =\frac 
	{2\int_{z^P_{ i, \rm min}}^{z^P_{ i, \rm max}}\mathrm{d}z^P_{G}\int_{z^P_{ i, \rm min}}^{z^P_{ i, \rm max}}\mathrm{d}z^P_g\int_{0}^{\infty}\mathrm{d}z_G W_L(z,z_G)p(z_G|z^P_G)p(z|z^P_g)S(z^P_G,z^P_g)n^P_i(z^P_G)n^P_i(z^P_g)}
	{\int_{z^P_{ i, \rm min}}^{z^P_{ i, \rm max}}\mathrm{d}z^P_{G}\int_{z^P_{ i, \rm min}}^{z^P_{ i, \rm max}}\mathrm{d}z^P_g\int_{0}^{\infty}\mathrm{d}z_G W_L(z,z_G)p(z_G|z^P_G)p(z|z^P_g)n^P_i(z^P_G)n^P_i(z^P_g)}  \label{eta}
\end{equation}

where $W_L$ is the lensing kernel, the superscript $P$ denotes photometric redshift, $p(z_G|z^P_G)$ is the photometric probability distribution function (PDF), $n^P_i$ is the photometric redshift distribution in the $i$th tomographic bin, and $S$ is the selection function defined further above in equation \ref{selection}. 

For this work we have assumed that the PDF is Gaussian of the form

\begin{equation}
    \label{eq:PDF}
    p(z|z^P) = \frac{1}{\sqrt{2\pi}\sigma_z\left(1+z\right)}\exp\left( -\frac{\left(z-z^P\right)^2}{2\left(\sigma_z\left(1+z\right)\right)^2}\right)
\end{equation}

In this Letter we have used $\sigma_z = 0.082$. With these tools in mind, we can move on to the separation of the correlation functions. In \cite{Zhang:2008pw} the work is done in $\ell$ space, but here we will instead focus on real space, to do this we define a constant $\hat{Q}_i$ as the average of $Q_i\left(\ell\right)$ over a reasonable range of $\ell$. With this, we can then perform a Hankel transform to real space as outlined in \cite{Joudaki:2017zdt}:
\begin{equation}
\label{eq:Hankel}
    w^{\lbrace Gg,Ig\rbrace}\left(\theta\right) = \frac{1}{2\pi} \int \mathrm{d}\ell \, \ell C_{\lbrace Gg,Ig\rbrace}\left(\ell\right) J_2 \left( \ell \theta\right),
\end{equation}
where $J_2$ is the second-order Bessel function, similarly for $w^{\gamma g}$ and $w^{\gamma g}|_S$. Recalling that $\hat{Q}_i$ is now a constant, and hence is not affected by the transform, we rewrite the expressions in real angular space.

Finally using the fiducial cosmology\footnote{We use the KiDS-450 fiducial cosmology obtained from \cite{2017MNRAS.465.1454H}} we switch from angular separation to perpendicular separation, $r_p$, to write the equations as a function of $r_p$ as follows:  
\begin{eqnarray}
    w^{Ig}\left(r_p\right) &= \frac{\left.w^{\gamma g}\right|_S \left(r_p\right) - \hat{Q}_i w^{\gamma g}\left(r_p\right)}{1-\hat{Q}_i}\\
    w^{Gg}\left(r_p\right) &= \frac{ w^{\gamma g}\left(r_p \right)-\left.w^{\gamma g}\right|_S \left(r_p\right)}{1-\hat{Q}_i}
    \label{split}
\end{eqnarray}

where the terms here can be obtained via the Treecorr code from \cite{Jarvis:2003wq}, and the $\hat{Q}_i$ can be obtained separately for each bin as shown in the right panel of Fig \ref{fig:IAmodel}. 
Similarly we can, as long as the galaxy bias $b_i$ is approximately constant (which we found it to be here), transform \eqref{eq:scaling} to write the scaling relation
\begin{equation}
\label{eq:scalingreal}
    w^{IG}_{ij}\left(r_p\right) \approx \frac{W_{ij}\Delta_i}{b_i} w^{Ig}_{ii}\left(r_p\right).
\end{equation}

Finally, we have generated a theoretical model merely for comparisons. This is done using the fiducial cosmology estimated from KiDS-450 \cite{2017MNRAS.465.1454H}, along with the amplitude $A_\text{IA}$ of the Intrinsic alignment for the tidal alignment model (\cite{2007NJPh....9..444B}) that is used as the default model for KiDS-450, as detailed in section 2.6 of \cite{2020MNRAS.495.3900Y}.

\section{Detection of GI-type intrinsic alignment Using the self-calibration method. } \label{sec:GI}
We have designed two pipelines for the separation of the Ig and Gg signals, as a way to cross-validate our results. Pipeline-1 was designed as a tool using AstroPy  \citep{2013A&A...558A..33A,2018AJ....156..123A} and SciPy \citep{scipy} to calculate the integrals needed to obtain the $Q$s as given in Equation \eqref{eq:Q}. The correlations shown in the right panel of Figure \ref{fig:IG-signal} for this pipeline are calculated using version 3 of Treecorr \citep{Jarvis:2003wq}, with jackknife regions obtained using the tiles from KiDS-450 \citep{2017MNRAS.465.1454H}. A first separation of the Ig correlation using the self-calibration method was obtained with Pipeline-1 \citep{Yao_dissertation} in the third redshift bin of KiDS-450 data. 

Pipeline-2 was designed to be adaptable and compatible with future surveys as well. For the $Q$ calculation, we use the Core Cosmology library (CCL; \citep{Chisari:2018vrw}) for calculating the linear power-spectra needed in Equations \eqref{eq:CGg} and \eqref{eq:CGgS}. Single and double integrals were calculated using SciPy \citep{scipy}. To solve triple integrals we use Monte Carlo integration to obtain a reliable result in a reasonable time, using the SciKit-Monaco code \footnote{https://pypi.org/project/scikit-monaco/}. For the correlations we use Treecorr 4.1 \citep{Jarvis:2003wq} with jackknife regions obtained via the newly implemented internal algorithm of TreeCorr. For the detection we used a fixed-size random catalog containing $10^8$ objects generated with the help of Healpix\_Util \footnote{https://github.com/esheldon/healpix\_util}. For the line-of-sight direction, we generate the distribution such that it corresponds to the true redshift distribution we estimated from the PDF model of Equation \eqref{eq:PDF}. We generate the true redshift distribution by stacking our PDF model; a longer discussion of this methodology in connection to the KiDS-450 dataset is given in \cite{2020MNRAS.495.3900Y}. This gets turned into a cumulative distribution function, which we can use to generate the fixed number of random redshifts we need. \textcolor{black}{More elaborate schemes for the randoms catalogs may be needed for surveys like Dark Energy Survey (DES) and Rubin Observatory LSST that are not limited to single epochs for each band in order to address some of the issues described in, for example, \cite{Leistedt:2015kka}}.

A measurement of the Ig correlation was obtained with a significance of 3.65$\sigma$ in the third bin of the KiDS-450 dataset, using the second pipeline.
The two independent pipelines have been used to extract the Ig correlations separately. 
The errors shown in Figure \ref{fig:IG-signal} are jackknife estimated. For Pipeline-2, the jackknife covariance for $w^{Ig,Gg}_{ii}$ is obtained by combining the covariances of $w^{\gamma g}$ and $\left.w^{\gamma g}\right|_S$ including the cross-covariance between these two correlations.

We used the scaling relation \eqref{eq:scalingreal} to obtain a \textcolor{black}{ 3.51$\sigma$ }measurement of the IG signal by cross-correlating bin 3 and bin 4. 
The IG result is depicted in the left panel of Figure \ref{fig:IG-signal}.
For this, we used Equation \eqref{eq:scalingreal}, where the scaling coefficient can be calculated for each separation, with the only thing varying being the galaxy bias $(b_i)$ which is almost constant across the bin. We used error propagation as described in the appendix of  \cite{Yao:2017dnt} to obtain the errors on IG correlation. 

It was found in \cite{2017MNRAS.465.1454H} and \cite{2020MNRAS.495.3900Y} that the bias caused by photo-$z$ outliers decreases noticeably from the two low-redshift bins (1 and 2) to the high-redshift bins (3 and 4) (see, for example. Section 3.4 in \cite{2020MNRAS.495.3900Y}). This is also reflected in our Figure \ref{fig:IAmodel} (right panel) for the $Q$ parameter curves that show clearly a good approximation to a constant for bins 3 and 4 compared to bins 1 and 2. This leads us to focus on bins 3 and 4. Additionally, the self-calibration method is designed to work in the case where the bins are such that $i<j$ due to the geometry of the IG type of IA requiring the intrinsically aligned galaxy to be in front of the sheared galaxy; see, e.g. \cite{Zhang:2008pw}. This leaves us with the combination of bins 3-4 for the self-calibration scaling relation.

\textcolor{black}{Next, we discuss the effect of nonlinear galaxy bias on our results. The bias enters the self-calibration calculations in two places. First it enters into the calculation of $Q_i$ using Equation~\eqref{eq:Q} where it arises in the integrals for the numerator, Equation~\eqref{eq:CGgS}, and the denominator, Equation~\eqref{eq:CGg}, of the ratio. Since it arises in both and as long as it remains nearly constant with respect to the distance $\chi$ (or redshift) within the bin, it can be pulled out of both integrals and will thereby cancel out. Our calculation of the galaxy bias for the four redshift bins finds that this is indeed a reasonable assumption. The second place where the galaxy bias arises is in the scaling relation, Equation~\eqref{eq:scaling}. We note that the study \cite{Yao:2018pgk} expanded this scaling relation to include nonlinear bias to second order and found that the denominator in this scaling relation is changed by an additional second-order bias term proportional to the bispectrum that is zero if non-Gaussianity in the density field is negligible and ignored. Our results are thus robust within the assumption of negligible non-Gaussianity in the density field, which is reasonable for the scope of this work. 
Moreover, it was also found in \cite{Zhang:2008pw} that the error on the galaxy bias in the scaling relation \eqref{eq:scaling} is subdominant to the error on the intrinsic alignment galaxy density correlation. We also find this to be the case in our calculation   of the linear bias using $w^{gg}_{ii}(r) \approx b_i^2 w^{mm}_{ii}(r)$, where $w^{gg}_{ii}$ is the galaxy density--galaxy density correlation from our samples and $w^{mm}_{ii}$ is the matter--matter correlation for the bin that we calculate theoretically using the true redshift distribution and the CCL \citep{Chisari:2018vrw}. With this, we obtain and use for each data point an estimated galaxy bias and its errors that we propagate throughout. We also included the numerical errors obtained on the $W_{ij}$ and $\Delta_i$ quantities.}

For galaxy selection, we use the cuts of the KiDS-450 data release  \citep{2017MNRAS.465.1454H}. We also explored making cuts based on color to focus for example on red galaxies,  We did explored making cuts based on color like focusing on red galaxies: however, these cuts reduced the galaxy samples too much and did not allow one to obtain sufficient statistics. Hopefully with other incoming larger surveys we can explore those kinds of selections and their effects on the self-calibration.

 We find here a negative IG intrinsic alignment signal using the self-calibration method in the third and fourth bins. This is in agreement with the negative IA amplitude of $|A_\text{IA}|=1.1$ found by KIDS450 team in \cite{2017MNRAS.465.1454H} from using the marginalization method and the linear tidal alignment model  \citep{Catelan2001,Hirata2004,2007NJPh....9..444B}.  In brief, this model is physically motivated by the concept that large-scale correlations between intrinsic ellipticities of galaxies are related linearly to perturbations in the primordial gravitational tidal field where these galaxies formed, leading to galaxy linear alignment with large-scale structure.
 The IG type of IA comes from the fact that a structure of matter aligns radially its close galaxies and shears tangentially background galaxies creating a correleation between these two as shown in Figure 1 (left panel). Our results are also consistent with the further findings of the KiDS+Gamma results of \cite{2019A&A...624A..30J} where a negative amplitude also was found using the same model. We do not use any IA model in the self-calibration, but for comparisons we plotted the theory curves from this linear tidal alignment model with the IA amplitude as found in the in KIDS450 paper \citep{2017MNRAS.465.1454H}. Some overall consistency is good to find, but it is not surprising that some discrepancies would be present -- IA modeling is an active area of work and self-calibration offers the opportunity to test such models once the signal is extracted. Although we are limited here in pursuing such a task, we expect in future work to pursue such an endeavor using more and better quality data from ongoing and planned surveys to constrain models using this approach. This will also allow one to compare to models like the Halo model introduced in \cite{Piras:2017rxk}, or the higher-order nonlinear model of \cite{Blazek:2017wbz}.

\section{Conclusion.} \label{sec:conclusion}
A first detection of intrinsic shape--gravitational shear (IG) and the intrinsic shape--galaxy density (Ig) in a photometric redshift survey using the self-calibration method is reported. The IG cross-correlation between the third and fourth bins of the KiDS-450 dataset is measured with a significance of \textcolor{black}{3.51$\sigma$}. The Ig correlation is measured in the third bin with a \textcolor{black}{3.65$\sigma$} significance. The negative IA signal we find here from the self-calibration method in the third and fourth bins, using the BPZ-determined best redshift estimate of \cite{2017MNRAS.465.1454H} is in agreement with the negative sign IA amplitude found there using the marginalization approach and the BPZ method and also is consistent with the results of \cite{Johnston:2018nfi} the combined GAMA and KiDS samples. We focused here on the third and fourth bins in KiDS-450 in view of their better quality photo-$z$. The self-calibration method has the advantage of not requiring the specification of an intrinsic alignment model. On the contrary, when an IA signal is extracted, it can be used to test and validate such models. It is worth noting that two independent pipelines have been used to derive the results for Ig correlations and were found to be in good agreement. These results also confirm that the self-calibration method works and show that it provides a means of extracting and mitigating intrinsic alignment signals from important future photometric surveys such as Rubin Observatory LSST, Euclid, and WFIRST.

\acknowledgments
We thank Anish Agashe, Hendrik Hildebrandt, Mike Jarvis, Lindsay King, Huanyuan Shan, Michael A. Troxel, and Haojie Xu for useful discussions. 
E.P. thanks his wife Shelbi Parker for proofreading the manuscript.
 M.I. acknowledges that this material is based upon work supported in part by the U.S. National Science Foundation under grant AST-1517768 and the U.S. Department of Energy, Office of Science, under Award Number DE-SC0019206. The authors acknowledge the Texas Advanced Computing Center (TACC) at The University of Texas for providing HPC resources that have contributed to the research results reported within this Letter. URL: http://www.tacc.utexas.edu. Based on (KiDS) data products from observations made with ESO Telescopes at the La Silla Paranal Observatory under program IDs 177.A-3016, 177.A-3017, and 177.A-3018. 

%

\vspace{5mm}


\software{astropy \citep{2013A&A...558A..33A},
          Treecorr \citep{Jarvis:2003wq}, 
          CCL \citep{Chisari:2018vrw}
          Scipy \citep{scipy},
          Scikit Monarco \url{https://pypi.org/project/scikit-monaco/},
          Healpix-util \url{https://github.com/esheldon/healpix\_util}
          }

\bibliography{bibliography}



\end{document}